\documentclass{mem}
\usepackage{natbib}\usepackage{txfonts}\usepackage{balance}
\usepackage{flushend}
\usepackage{graphicx}
\usepackage[breaklinks,pdftex]{hyperref}
\idline{75}{282}
\begin{document}
\def\teff{$T\rm_{eff }$}
\def\kms{$\mathrm {km s}^{-1}$}

\title{
Observation of Astrophysical Sources with SST-1M Telescopes - First Results
}

\author{
J. \,Jury\v{s}ek\inst{1} 
\and 
T. \,Tavernier\inst{1} 
\and 
V. \,Novotn\'y\inst{1, 2} 
for the SST-1M Collaboration
}

\institute{
FZU - Institute of Physics of the Czech Academy of Sciences, Na Slovance 1999/2, Prague 8, Czech Republic, \email{jurysek@fzu.cz}
\and
Charles University, Faculty of Mathematics and Physics, Institute of Particle and Nuclear Physics, Prague, Czech Republic
\\
}

\authorrunning{Jury\v{s}ek}
\titlerunning{SST-1M: first results}

\date{Received: XX-XX-XXXX (Day-Month-Year); Accepted: XX-XX-XXXX (Day-Month-Year)}

\abstract{
The Single-Mirror Small Size Cherenkov Telescope (SST-1M) was developed by a consortium of institutes in Switzerland, Poland, and the Czech Republic. The SST-1M design is based on the Davies-Cotton concept, featuring a 4-meter mirror and an innovative SiPM-based camera. It is most sensitive to gamma rays in the TeV and multi-TeV energy bands. Since 2022, two SST-1M prototypes have been commissioned at the Ond\v{r}ejov Observatory in the Czech Republic, where their performance in both mono and stereo observation modes is being tested. During the commissioning phase, several galactic and extragalactic gamma-ray sources have been observed, resulting in multiple detections. In this contribution, we present preliminary results from this observation campaign.
}
\maketitle{}

\section{Introduction}

The Single-Mirror Small Size Cherenkov Telescope (SST-1M) was initially developed as part of an array of over 50 telescopes designed to extend the sensitivity of the Cherenkov Telescope Array Observatory (CTAO) for detecting gamma rays with energies greater than approximately 1 TeV \citep[e.g.][]{2019ICRC...36..694H}. The 4-m mirror of the telescope consists of 18 hexagonal facets organized into a spherical dish following the Davies-Cotton design. It ensures a good off-axis point spread function, crucial when aiming for large field-of-view (FoV) observations \citep{sst1m_hw_paper}. The camera consists of 1296 Silicon Photomultiplier pixels, and a fully digital trigger and readout system, the Digicam \citep{2017EPJC...77...47H}, allowing for high Night Sky Background operation, which increases the duty cycle of the observations \citep{nagai_sipm_2019}.

In 2022, two SST-1M telescope prototypes were installed in Ond\v{r}ejov, Czech Republic, at 510 m.a.s.l. to test the telescopes until their final site is decided. The telescopes are separated by 152.5 meters, and their timestamps are synchronized with nanosecond precision using the White Rabbit server, which enables stereoscopic observations and triggering managed with the CTA Software Array Trigger (SWAT). Currently, both telescopes are undergoing commissioning focused on the telescope operation, calibration, and observation of galactic and extragalactic gamma-ray sources.

\begin{figure}[!t]
\resizebox{\hsize}{!}{\includegraphics[width=.90\textwidth]{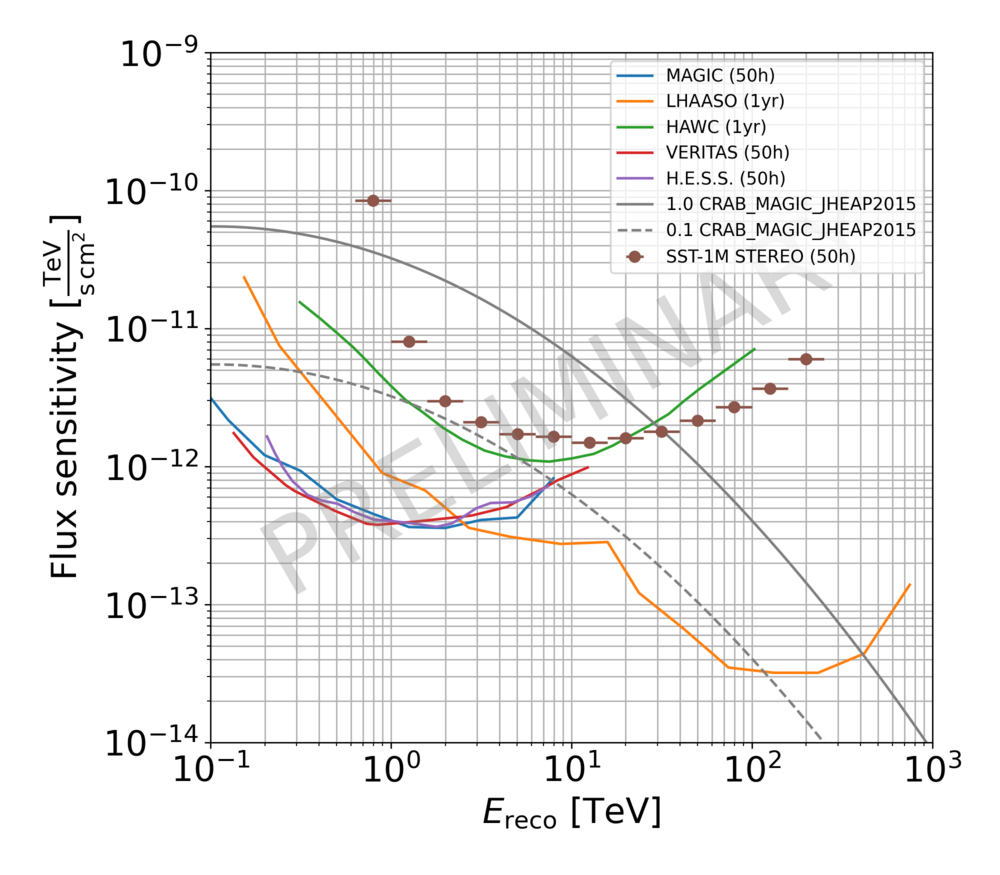}}
\caption{
\footnotesize
SST-1M stereo sensitivity at low altitude, compared with major existing IACTs and Water Cherenkov experiments.}
\label{fig.sensitivity}
\end{figure}

Despite the difficult atmospheric conditions in Ond\v{r}ejov limiting the number of clear nights, the physics performance turned sufficient for interesting science cases. The angular and energy resolution in stereo reaches $\approx0.07^\circ$ and $\approx10\%$, respectively \citep{2024icrc.confE.592J}. Figure~\ref{fig.sensitivity} shows the stereo sensitivity of the SST-1M observatory compared with other existing major ground-based gamma-ray experiments. One can note that above few tens of TeVs SST-1M overcomes all existing Imaging Atmospheric Cherenkov Telescopes (IACTs), at those energies limited mostly by their small FoV.

\section{Results}

In this section, we discuss the detection of three sources observed with SST-1M in stereo during the observation campaign between autumn 2023 and summer 2024. The observations were performed in wobble mode \citep{1994APh.....2..137F} with an offset ranging from $0.7^\circ$ to $1.4^\circ$. The calibration of raw data \citep{2024arXiv240918639T} and event reconstruction is performed with \texttt{sst1mpipe}\footnote{\url{https://github.com/SST-1M-collaboration/sst1mpipe}}, a pipeline developed particularly for SST-1M stereoscopic data analysis \citep{jurysek_2024_10852981, 2024icrc.confE.592J}. 

\subsection{Crab Nebula}

Being the brightest persistent, almost point-like VHE gamma-ray source in the sky, the Crab Nebula is considered a standard candle of gamma-ray astronomy. The stereoscopic observations of the Crab Nebula were conducted between October 2023 and March 2024, which resulted in 23 hours of good stereo data with zenith angles between $25^\circ$ and $45^\circ$ after quality cuts. $5\sigma$ detection of the Crab Nebula in stereo was reached in 1.5 hours, which meets the expectations based on MC simulations \citep{2024icrc.confE.592J}. 

\begin{figure*}[!t]
\centering
\begin{tabular}{cc}
\includegraphics[width=.44\textwidth]{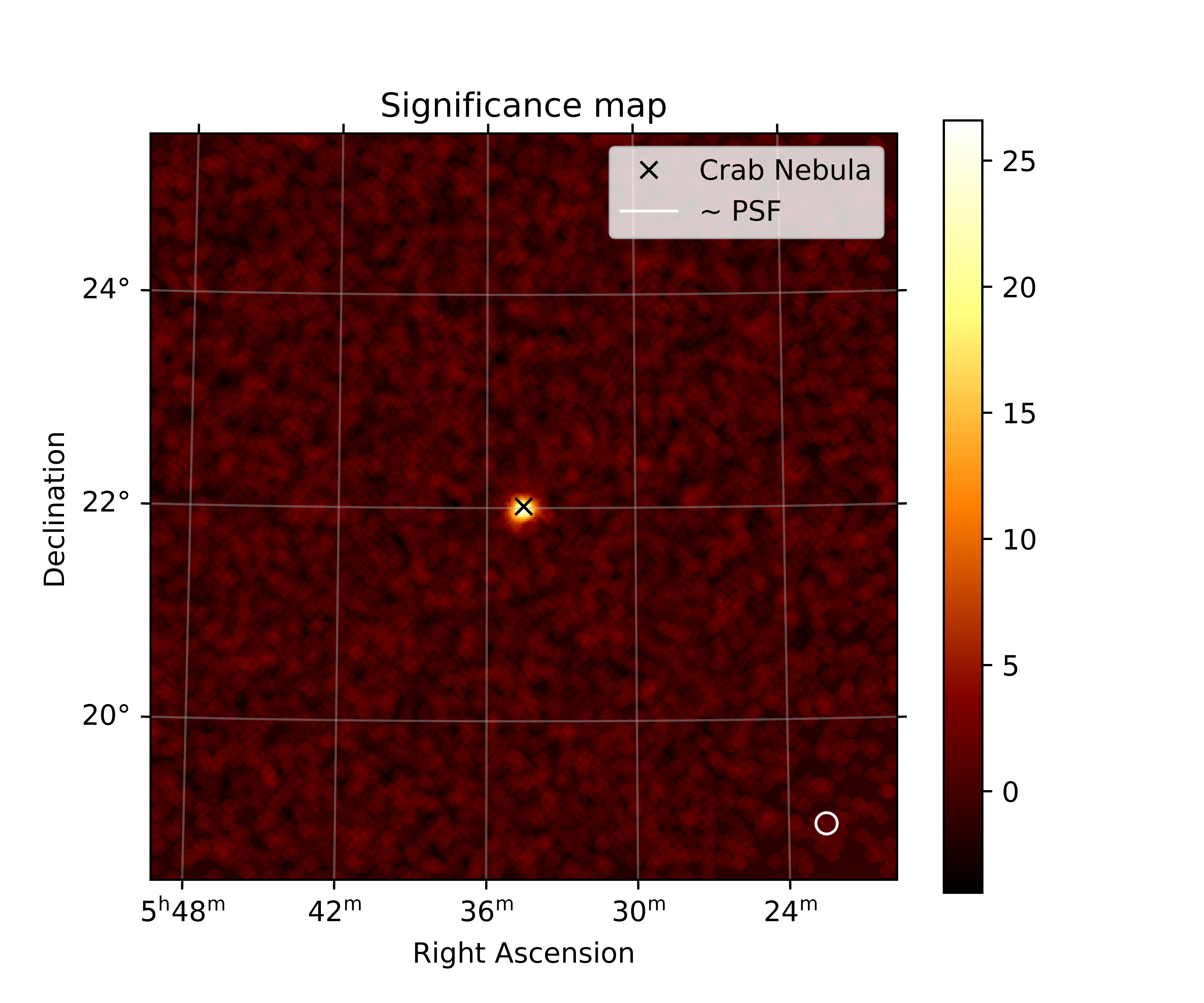} & \includegraphics[width=.44\textwidth]{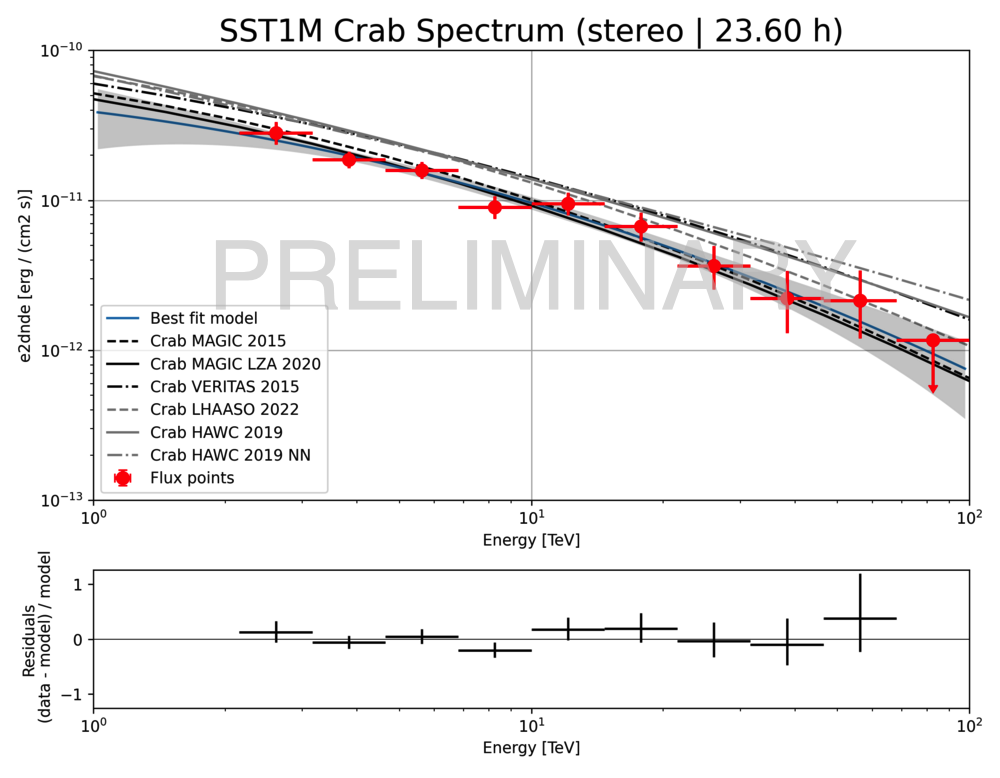}
\end{tabular}
\caption{The Crab Nebula measured with SST-1M in stereo. \textit{Left:} Significance map. \textit{Right:} SED compared with the results of other experiments.}
\label{fig.crab}
\end{figure*}

Left Figure~\ref{fig.crab} shows $7^\circ \times 7^\circ$ significance map of the region around the Crab Nebula obtained with the ring background method. The map is plotted for the integrated energy range $1 - 100$ TeV and the convolution radius of $0.1^\circ$, corresponding to the averaged SST-1M stereo PSF. One may notice an outstanding background homogeneity ($\mu=-0.11$ and $\sigma=1.03$) on the scale of a few degrees, which promises outstanding SST-1M capabilities for the observation of extended galactic sources.

Figure~\ref{fig.crab} - right shows a preliminary Spectral Energy Distribution (SED) of the excess events in the signal region after subtraction of the reflected background. To limit the systematics induced by a possible MC-data disagreement, we rejected all events in the reconstructed energy bins with effective area $< 5\%$ of the maximum and with the energy bias $> 10\%$. Assuming the Log-parabola spectral shape for the differential flux $d\phi / dE$ in the form of $d\phi / dE = \phi_0 (E / E_0)^{-\alpha - \beta \log{E/E_0}}$, where the reference energy $E_0 = 6.31$ TeV was fixed on the value where all parameters are the least correlated. The best-fitting parameters are $\alpha=2.76\pm0.11$, $\beta=0.11\pm0.09$ and the flux normalization $\phi_0=(2.19\pm0.17) \times 10^{-13} \, \mathrm{cm^{-2} s^{-1} TeV^{-1}}$. We note that despite very good agreement with the results of other experiments, we neglect systematic uncertainties in this preliminary study. Detailed study of the SST-1M performance, MC/data agreement, Crab Nebula spectrum, and systematic uncertainties involved is the subject of a follow-up study.

\begin{figure*}[!t]
\centering
\begin{tabular}{cc}
\includegraphics[width=.95\textwidth]{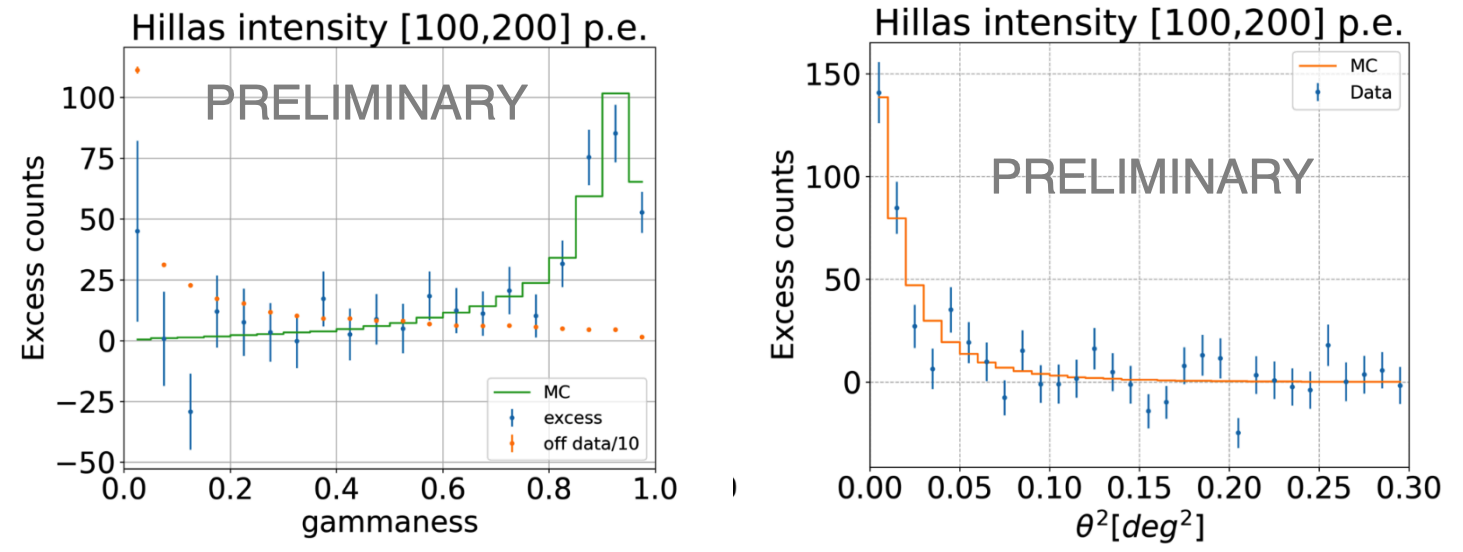}
\end{tabular}
\caption{Monte Carlo - data comparison for Crab Nebula excess events measured with SST-1M-1 in mono.}
\label{fig.mc_data}
\end{figure*}

The Crab Nebula dataset is crucial for the validation of the data analysis pipeline and agreement of the Monte Carlo model of the telescopes with data in order not to introduce unnecessary systematic biases. Figure~\ref{fig.mc_data} shows distributions of the Crab Nebula excess events measured with SST-1M-1 in mono, compared with point-like gammas Monte Carlo re-weighted on the Crab Nebula spectrum \citep{ALEKSIC201676}. The overall agreement demonstrates a good understanding of the telescope. 

\subsection{Mrk 421}

One of the SST-1M observatory's science objectives is monitoring of nearby blazars at TeV energies. We conducted mid-term (January - May 2024) monitoring of Mrk 421, which became the first extragalactic source detected with SST-1M in stereo. We found the source in a low state for almost entire period, except for significant brightening detected on March 13, when Mrk 421 flux reached the level of Crab Nebula, leading to 6.8$\sigma$ detection in 3.3 hours \citep{tavernier_atel, 2024arXiv240918587T}. The increased flux level was confirmed on March 17 with an additional 5 hours of observation \citep{2024arXiv240918587T}.

\begin{figure}[!t]
\resizebox{\hsize}{!}{\includegraphics[clip=true, width=.9\textwidth]{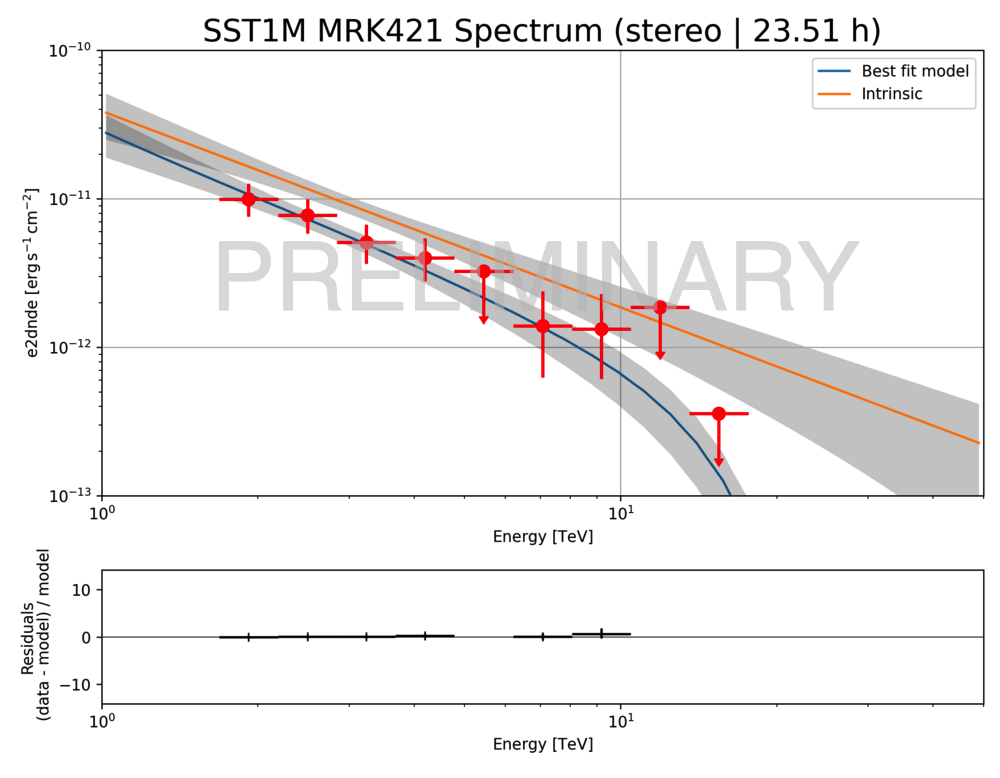}}
\caption{
\footnotesize
SED of Mrk 421 integrated over the entire period of observation (January - May 2024).}
\label{fig.mrk421}
\end{figure}

After selecting runs not affected by bad weather conditions, we ended up with 23.5 hours of good-quality data. We performed 1D spectral analysis of the stacked dataset for the entire observation period, in the range of reconstructed energies between 1 and 50 TeV. Figure~\ref{fig.mrk421} shows preliminary SED assuming Power-Law spectral shape for the intrinsic spectrum. The Extragalactic Background Light absorption was taken into account \citep{2011MNRAS.410.2556D}. The SED shows no spectral curvature ($\Delta$TS=0.03 for intrinsic ECPL spectral model over PL) and a steep spectral index of $3.22\pm0.3$. It can both be most likely attributed to the fact that SST-1M with an energy threshold higher than 1 TeV probes the energies above the spectral cutoff ($E_\mathrm{cutoff} = 5.1$ TeV in the HAWC long-term SED \citep{2022ApJ...929..125A}). Using ECPL spectral model with $E_\mathrm{cutoff}$ fixed on the HAWC reported value, the spectral index of our dataset is $2.6\pm0.3$, which is compatible with \citet{2022ApJ...929..125A} results ($2.26\pm0.12$).

\subsection{VER 2019+368}

The vicinity of the X-ray "Dragonfly" Pulsar Wind Nebula is a complex region with multiple HE to UHE gamma-ray point-like and diffuse sources accompanied with their multi-wavelength counterparts. VHE gamma-ray emission was discovered by MILAGRO \citep{Abdo_2012}, later resolved into more sources by VERITAS \citep{2014ApJ...788...78A, Abeysekara_2018}. Recently, the LHAASO observatory detected photons with energies up to 270 TeV \citep{2024ApJS..271...25C}, making the region a promising candidate for accelerating charged particles up to PeV energies. The core VHE emitting region around VER 2019+368 shows asymmetric extension ($0.34^\circ \times 0.13^\circ$ \citep{2014ApJ...788...78A}) and hints at energy-dependent morphology. 

\begin{figure*}[!t]
\centering
\begin{tabular}{cc}
\includegraphics[width=.44\textwidth]{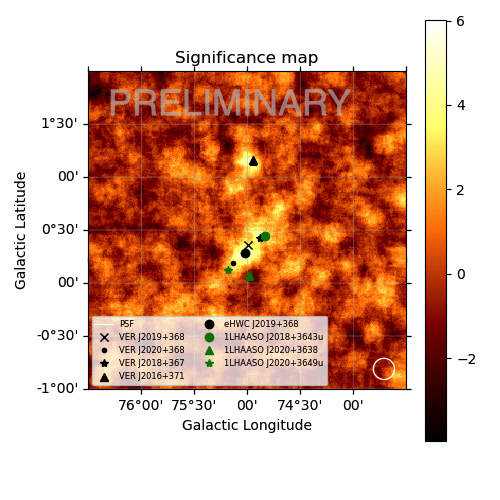} & \includegraphics[width=.44\textwidth]{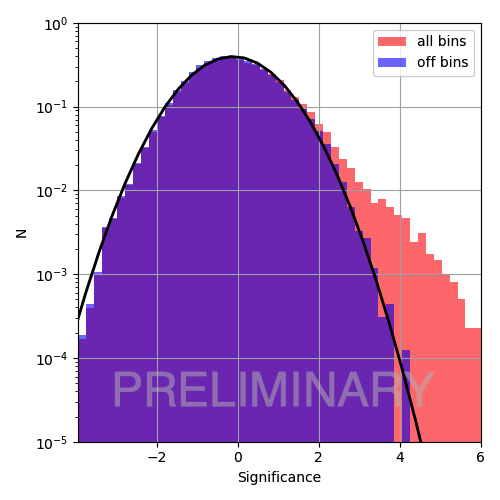}
\end{tabular}
\caption{\textit{Left:} Significance map of the VER 2019+368 region. \textit{Right:} Distribution of local significances for all positions in the skymap (red), and the background (blue).}
\label{fig.dragonfly}
\end{figure*}

The first observation campaign conducted with SST-1M occurred from April to August 2024, collecting 59 hours of good-quality stereo data between $5^\circ$ and $60^\circ$ zenith angles. Figure~\ref{fig.dragonfly} - left shows $3^\circ \times 3^\circ$ significance map of the Dragonfly region obtained with the ring background method. Regions within a radius of $0.3^\circ$ from all known nearby VHE sources were removed for background estimation. The energy range for the map is $1-300$ TeV and the convolution radius of $0.1^\circ$. Figure~\ref{fig.dragonfly} - right shows the distribution of the local significance for pixels in the skymap. The significance distribution for the background pixels can be described as Gaussian with $\mu=-0.12$ and $\sigma=1.01$. One can note that the excess regions correlate well with the position of known VHE sources and both VER 2019+368 and CTB87 regions are clearly resolved (compare with Fig. 5 in \citet{Abeysekara_2018}). We also performed an independent 1D spectral analysis of the same dataset with a fixed signal region on the VERITAS reported position and the same size (R.A.=304.8458, DEC=36.7789, r=0.23 \citep{Abeysekara_2018}), testing for the presence of a source. We found that a source with a PL spectrum is preferred over no source hypothesis at $6.9\sigma$ (pre-trial). Detailed study of the Dragonfly region SED and its morphology is the subject of a follow-up paper.

\section{Conclusions}
The first astrophysical observations during commissioning in Ond\v{r}ejov in both mono and stereo modes prove that the SST-1M telescopes meet the expected performance. The detection of the extended region of VER 2019+368 demonstrates the SST-1M capabilities for the detection of extended sources, and together with the sensitivity to the most energetic gamma rays and good angular resolution, it makes the instrument an ideal tool for morphological studies of extended galactic PeVatron candidates. 



\bibliographystyle{aa}
\footnotesize
\bibliography{bibliography}

\begin{acknowledgements}
\scriptsize
The work is financed by the D´epartment de Physique Nucl´eaire et Corpusculare, Faculty de Sciences of the University of Geneva, 1205 Geneve, and the construction of the SST-1M cameras was also supported by The Foundation Ernest Boninchi,1246 Corsier-CH, and the Swiss National Foundation (grants 166913, 154221, 150779, 143830). Funding by the Polish Ministry of Science and Higher Education under project DIR/WK/2017/2022/12-3 and 2024/WK/03 are gratefully acknowledged. The Czech partner institutions acknowledge support of the infrastructure and research projects by Ministry of Education, Youth and Sports of the Czech Republic and regional funds of the European Union, MEYS LM2023047 and EU/MEYS CZ.02.01.01/00/22 008/0004632, Czech Science Foundation, GACR 23-05827S, and Co-funded by the European Union (Physics for Future – Grant Agreement No. 101081515).
\end{acknowledgements}

\end{document}